# Efficient Extraction of Hot Carriers in Perovskite Quantum Dot through Building State Coupled Complex

Yusheng Li[a]†, Junke Jiang[b,f]†, Dandan Wang*[a], Dong Liu[a], Shota Yajima[a], Hua Li[a], Akihito Fuchimoto[a], Hongshi Li[c], Guozheng Shi[a], Shuzi Hayase[a,d], Shuxia Tao[b], Jiangjian Shi[e], Qingbo Meng*[e], Chao Ding*[a], Qing Shen*[a,d]

[a.] *Faculty of Informatics and Engineering, The University of Electro-Communications, 1-5-1 Chofugaoka, Chofu, Tokyo 182-8585, Japan.*

[b.] *Materials Simulation and Modelling, Department of Applied Physics, Eindhoven University of Technology, 5600 MB Eindhoven, The Netherlands.*

[c.] *Institute of New Energy Materials Chemistry, School of Materials Science and Engineering, Nankai University, TongYan street 38, Jinnan District, Tianjin 300350, China*

[d.] *CREST, Japan Science and Technology Agency (JST), 4-1-8 Honcho, Kawaguchi, Saitama 332-0012, Japan*

[e.] *Key Laboratory for Renewable Energy, Beijing Key Laboratory for New Energy Materials and Devices, Institute of Physics, Chinese Academy of Sciences, Beijing 100190, P. R. China*

[f.] *Present Address: Université Rennes, ENSCR, CNRS, ISCR-UMR 6226, F-35000 Rennes, France*

† These authors contributed equally *Corresponding authors: qbmeng@iphy.ac.cn, wdd2395360261@gmail.com, ding@jupiter.pc.uec.ac.jp, shen@pc.uec.ac.jp (Lead Contact)

# Abstract

Utilizing hot carriers is the crucial approach for solar cell to exceed the thermodynamic detailed balance limit, yet effective extraction of hot carriers in absorber materials via most commonly used semiconductor acceptors has been a challenge in both materials and photophysics research for many years. Herein, we build series of $CsPbI_3$ quantum dot and fullerene derivative systems to explore the decisive factors of this process and have for the first time realized efficient hot carrier extraction in these systems (maximum extraction efficiency ~ 84%). We find building the systems as state-coupled complexes creates new carrier transport channels at about 0.22 eV above $CsPbI_3$ quantum dot bandgap, which facilitates highly efficient HC extraction. Our research directly visualizes the inner connection of molecule interaction and ultrafast hot carrier extraction. The knowledge and strategy gained here are of universal meaning, taking an important step forward true hot carrier photovoltaics.

## Introduction

In all photovoltaic devices, efficient extraction of carriers from photon absorber materials is directly linked to overall device efficiency since the photogenerated carriers ultimately produce electricity through charge transport, extraction, and collection processes[1-3]. Semiconductor photovoltaic devices that only utilize the band-edge carries in absorber to produce electrical power, is limited to Shockley-Queisser limit efficiency (~33%)[4]. Much higher efficiencies, theoretically as high as 66%, are possible if the photogenerated hot carriers (HCs) can be efficiently extracted by suitable charge-selective materials before losing their excessive energy[5].

To realize extraction of HCs, slow HC cooling in absorber materials and effective HC transfer at interface are two indispensable parts for competing with the energy loss processes of HCs[6-8]. Extraction of HCs in conventional semiconductor absorber materials is extremely limited by the ultrafast HC cooling processes[7, 9]. Recently widely studied halide perovskites are revealed with remarkable long-lived HCs, about two to three orders of magnitude longer than conventional semiconductors[10-15], due to the presence of hot-phonon bottlenecks[11, 13], acoustic-optical phonon up-conversion[14], band-filling[12] and Auger heating[13]. In principle, the HC diffusion length in perovskites is sufficient to allow HCs transport to the absorber surface and transfer to generally used semiconductor charge-selective materials.

Unfortunately, broadly used inorganic charge-selective materials, such as $TiO_2$, $SnO_2$ have been indicated to be difficult to extract HCs from perovskite materials due to their weak charge extraction ability and lacking coupling with perovskite[16, 17]. Fullerene derivatives (FDs) were well known for their excellent performance as electron acceptor in organic solar cells[18, 19]. Nowadays, such materials have also been regarded as star materials in perovskite photovoltaic devices and gained great successes as the electron-selective materials, interface modification materials and even additives due to their excellent electrical properties and good chemical and physical coupling with perovskites[20, 21]. Moreover, they are suggested with the potential as hot exciton acceptor. Their blending with polymer aggregates facilitates hot exciton dislocation and excited electronic state coupling, producing an overall increase of organic device performance[22-24]. However, no similar experimental evidence has been presented in perovskite and FD systems. Achieving effective HC extraction in perovskite and FD systems is of great benefit to impel perovskite photovoltaics to a new plateau.

In this work, we choose colloidal perovskite quantum dots (PQDs) as absorber material, as they provide larger surface area for coupling with FD molecules and the HCs in QDs are easier to transport to surface than bulk materials. We build series of the $CsPbI_3$ QDs and FDs systems to probe their molecule interaction dynamics and ultrafast HC dynamics. We firstly explicate a general strategy to quantify HC relaxion for evaluating HC extraction efficiency and use it into our research of HC extraction. The completely distinct HC extraction in the systems with FDs owing different function groups are directly visualized by transient absorption (TA) spectroscopy. Using density of function theory (DFT) calculation and Stern-Volmer analysis, we elucidate the different molecule interaction mechanisms in these systems and their decisive role on the HC extraction in these systems. We find the key for realizing efficient HC extraction in $CsPbI_3$ QDs and FDs systems lies on whether building state-coupled complex. The maximum extraction efficiency of 84% is finally achieved in these complexes by adjusting pump power. We show that efficient extraction of HCs from $CsPbI_3$ QD to broadly used acceptors is feasible with building appropriate chemical combination. Our research updates the perspective of extracting HCs in perovskite materials and move a major step towards true HC photovoltaics.

## Results and discussions

**Building CsPbI$_3$ QD and FD systems**

CsPbI$_3$ QDs were synthesized by previous reported two sources hot injection method in oleic acid (OA), oleylamine (OAm), and octadecene[25, 26]. The as-prepared CsPbI$_3$ QDs crystallized in the cubic phase (Figure S1), with an average diameter of 14.6 nm (Figure S2) and exhibited a photoluminescence quantum yield (PLQY) of ~95%, suggesting high crystalline quality. We synthesized CAFDs (PCBA and [Bias]PCBA) from the acid hydrolysis of PCBM and [Bias]PCBM respectively (synthetic details are given in supplementary methods) as semiconductor selective materials for CsPbI$_3$ QDs. For comparation, same molar quantities of PCBM and [Bias]PCBM (called CMFDs, i.e., common fullerene derivatives) were used as reference materials. The molecular structures are shown in Scheme 1 and relevant nuclear magnetic resonance spectroscopy (NMR), optical absorption, and energy level are shown in Figure S3-5. The FDs-adsorbed CsPbI$_3$ QDs dispersions (indicated hereafter as CsPbI$_3$-FD QDs) were obtained by mixing CsPbI$_3$ QDs with small quantity of FDs, and then purified by successive precipitation and centrifugation cycles (see experiments details in supplementary information). XRD (Figure S2) and UV-vis absorption spectra (ABS, Figure S6) indicate no explicit detrimental effects of FDs on the chemical structure of CsPbI$_3$ QDs and the absorption contribution from FDs is completely negligible in our experiments.

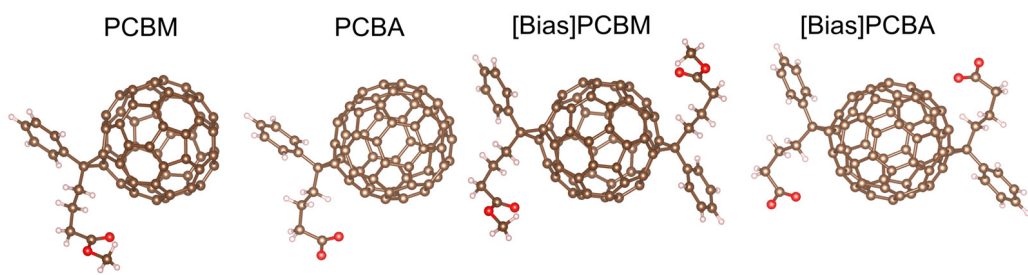

Scheme 1. Molecular structures of PCBM, PCBA, [Bias]PCBM, and [Bias]PCBA. We note both [Bias]PCBM and [Bias]PCBA are the mixture of isomers. Here, we only show one structure of the isomers.

**Hot carrier relaxion dynamics**

Pump power dependent HC cooling mechanisms complicate the evaluation of the HC relaxation time and determination of HC extraction efficiency[13, 27]. Building a convenient method with general applicability is still an imperative. In this part we will analyze the HC dynamics in $CsPbI_3$ QDs, then evaluate a reliable method for our HC extraction research. Figure 1a,b shows the pseudo color TA maps of $CsPbI_3$ QDs excited at 470 nm under a lower (16 μJ/cm$^2$) and higher (200 μJ/cm$^2$) pump fluence, corresponding to initial average generated exciton (e-h pair) numbers per QDs *<N>*~0.2 and *<N>*~2.3 respectively (see Figure S7). For both cases, immediately after photoexcitation, positive photo induced absorption (PIA) peak emerged below the bandgap with respect to bandgap remoralization (BGR) effect[13, 28]. Decay of PIA occurred once the HCs cooled down to occupy the states near band edge, simultaneously, band filling induced negative photobleaching (PB) signal with high energy tail gradually settled in near the bandgap and finally replaced the PIA signal. In lower intensity pump, global analysis of TA spectrum (Figure S8, evolution-associated spectra (EAS)) indicates PIA displays a consistent dynamic with the broadened PB. The relevant kinetical curve of this EAS component is shown in Figure 1c (dashed black line). In previous research, PIA decay has been used to evaluate HC lifetime[28]. The kinetic curve that represents PIA was chosen at 1.76 eV, where the contribution of PB signal is negligible. Along with the thermalization and relaxion of HCs, the ground state bleaching (GSB) gradually increases and reaches the maximum, which can be fitted with a convolution of single-exponential growth function and instrument response function (IRF) of TA measurement to yield a rise time. In some research, the HC lifetime was also estimated by monitoring this rise time[29, 30]. When the HCs totally cool down to be excited carriers at band edge, the broadened PB with high-energy tail will compress to the shape close to secant hyperbolic function or Gaussian function broadened GSB of exciton (Figure S8). Note that the weak positive absorption band (PIA2) existed at the higher energy side of the isosbestic point (IBP at 1.96 eV in Figure S8) on GSB shows small deviation accounted for excited carrier absorption at band edge[27]. In the presence of HCs, isosbestic point will shift to higher energy, at the meantime, the signal at the energy corresponding to the isosbestic point of GSB becomes negative PB, which means any new PB feature observed at this point can be attributed to the HC signal. In view of this, we propose to use the transient kinetic at this IBP to image the bleaching signal of HC.

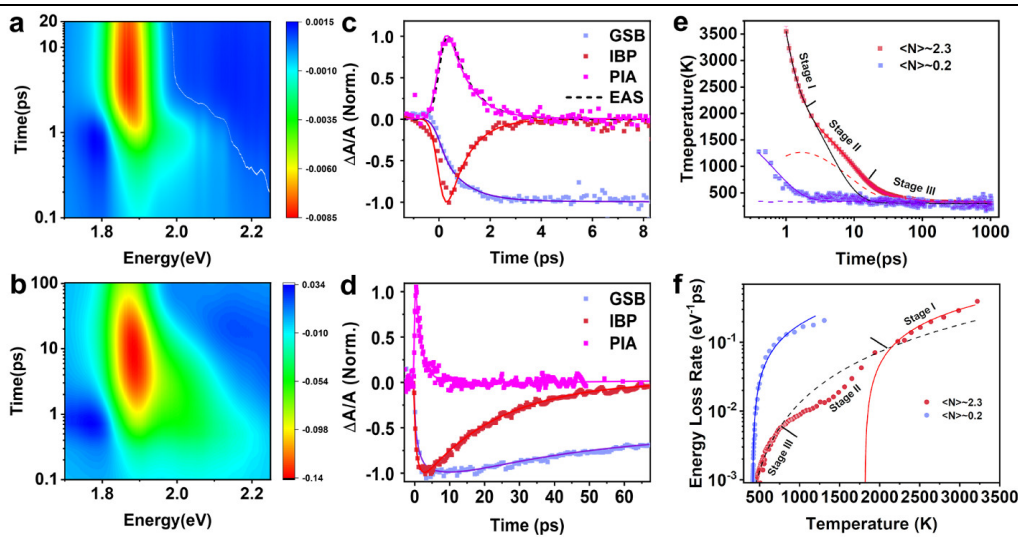

Figure 1. TA spectra and HC cooling dynamics of CsPbI$_3$ QDs. a-b, Pseudo color TA maps under a 470 nm pump for (a) low (16 μJ/cm$^2$, <N>~0.2) and (b) high (200 μJ/cm$^2$, <N>~2.3) pump fluence. c-d, Evolution of HC lifetime with kinetics of ground state bleaching (GSB), isosbestic point (IBP), photo induced absorption (PIA) and evolution-associated spectra (EAS) for (c) low and (d) high pump fluence. (e) Time-dependent HC temperature $T_c$. Solid red line and purple line represent the fitting curves with three temperature model (TTM). Dash red line and purple line are phonon temperatures. The solid black line is the HC temperature decay under higher pump fluence after removing the contributions of Auger reheating. (f) Energy loss rate as a function of HC temperature. Solid red and blue lines represent the numerical fits with LO–phonon model, while dashed black line represents the fit with Auger reheating model.

As shown in Figure 1c, the GSB formation, PIA decay, IBP decay and EAS decay at lower pump, are kinetically correlated with a similar lifetime ~0.84 ps, which means these methods tend to acquire a same HC relaxion time. However, these kinetical curves are no longer completely synchronized at high-fluence pump (Figure 1b,d). Obviously, the decay at IBP exhibits slower dynamic than the formation of GSB and the PIA decay but still keeps in consistent with the existence of high energy tail that reveals the energy distribution of HCs. Note that EAS misses in Figure 1d since it cannot provide a clear picture to depict the dynamic relevant components under the influence of more obvious band filling effect as shown in Figure S8. We ascribe the non-synchronous dynamics in these curves to the appearance of multi-particle effects including Auger recombination and concomitant Auger heating, in which Auger recombination losses cold carriers and displays as the intensity-dependent rapid shortening of the GSB dynamics (Figure S7), while Auger heating contributes to further retardation of HC relaxation and plays an important role in practical HC solar cells[13, 27].

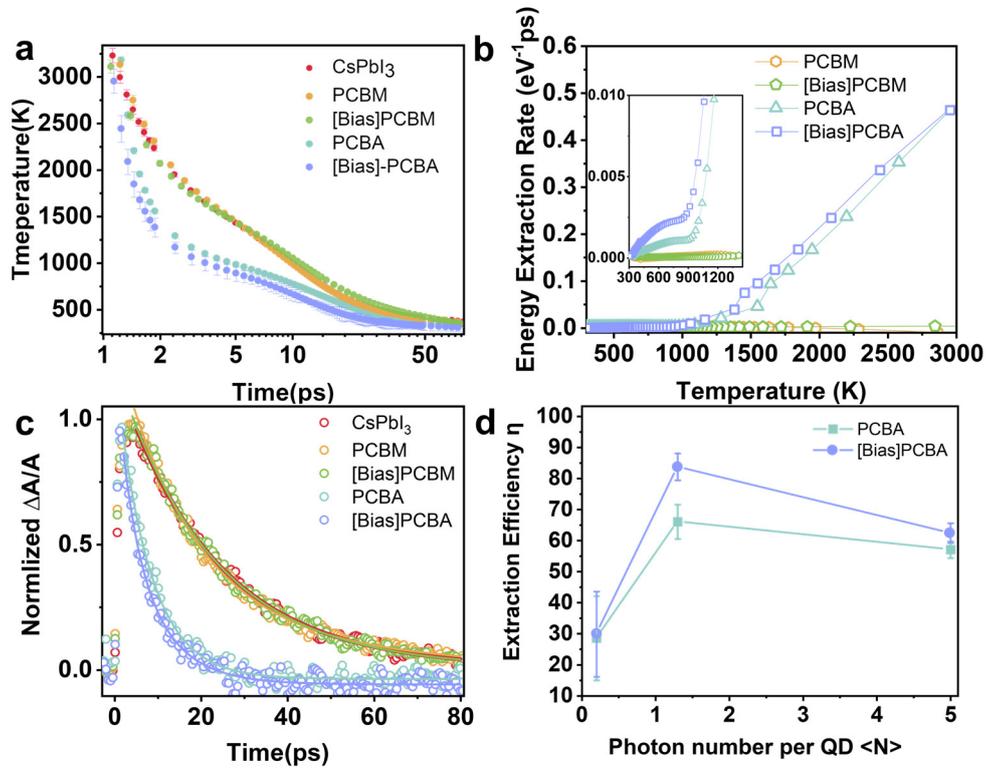

Figure 2. Hot carrier extraction of FDs. (a) Time-dependent HC temperature $T_c$. (b) Energy extraction rates as a function of HC temperature. Inset is zoom of (b) at low

carrier temperature. (c) IBP kinetics. (d) HC extraction efficiency with photon numbers per QD.

Intrinsic energy losses of CsPbI$_3$ QD in the processes of HC relaxion are further discussed via the analysis of time-dependent carrier temperature $T_c$ (Figure 1e). To ensure that all population of HCs reach quasi-equilibrium and occupy the states at near band edge forming Fermi-Dirac distribution, we analyzed the HC cooling dynamics after a 0.3~1 ps delay (depend on pump intensity). Average carrier temperature $T_c$ can be extracted by fitting the high-energy band tail of GSB[13, 30] according to $1/(1+\exp(E-E_f)/k_B T_c)$, where $E_f$ is quasi-Fermi energy, $k_B$ is Boltzmann constant (Figure S9). The initial excess energy ($\Delta E_{excess}$) of carriers stimulated with 2.64 eV (470 nm) pump energy ($E_{pump}$) relative to band edge is expected to be 0.36 eV corresponds to a highest effective temperature $T_c$ ~2700 K (e.g., *$\Delta E_{excess}=(E_{pump}-E_g)/2-E_g$ and $T_c$~$2\Delta E/3k_B$*, assuming electron and hole in perovskites have similar the effective mass and hold similar temperature with each other[11, 29, 31]). The drop of the effective temperature of carriers from initial pump to quasi-equilibrium are ascribed to carrier-carrier scattering. Subsequent temperature losses are numerically fitted with three temperature model (TTM)[32, 33], which provides a clear picture of various stages of HC relaxation by coupling Auger reheating into the rate equations of HCs and longitudinal optical (LO) phonons (supplementary Note 1). Under low pump fluence, the HC relaxion displays single exponential decay and takes around 0.76 ps (Figure 1e), which is similar to the dynamics of GSB formation, PIA decay, IBP kinetic in Figure 1c. The scattering interaction between carrier and longitudinal optical (LO) phonons contributes to the fast drop of HC temperature. Whereafter, the LO-phonons rapidly release obtained excess energy to lattice without bottleneck, see LO-phonon temperature ($T_p$, dash purple line in Figure 1e). The energy power loss rate $J_r$ (e.g.*-1.5$k_B T_c$*) in this regime can be fitted with LO-phonon interaction model (supplementary Note 2), the fitted LO–phonon decay time $\tau_{LO}$ (~0.51 ps) is consistent with previous reports[34].

Under high pump fluence, HC relaxation changes significantly compared with that at low-fluence pump. Specifically, the HC cooling process is not terminated within picosecond but prolonged to tens of picoseconds. In this case, the time dependent temperature can be divided into three stages. time dependent temperature can be divided into three stages. Stage I is dominated by LO-phonon interaction as same as temperature decay under low pump fluence. It can be demonstrated from the similar value of $\tau_{LO}$ (0.47 ps) fitted from the energy power loss rate at this stage. The difference of stage I with the temperature decay under low pump fluence lies on HCs eventually heated the LO-phonon to a higher temperature, as the dash red line in Figure 1e. After the LO-phonons attaching to peak temperature, the LO-phonons release energy to acoustic phonon or lattice with a slow rate probably due to hot phonon bottleneck effect. In addition, Auger reheating also contributes the slow HC relaxation in this stage, which can be found in the accelerated temperature decay after removing the Auger reheating effect (solid black line in Figure 1e). We think stage III is mainly on account of Auger reheating effect since the behavior of energy power loss rate follows quadratic dependence[34] on ($T_c$-$T_L$) remarkably well ($T_L$ is lattice temperature) and phonon bottleneck can be neglected due to $T_p$ is close to $T_L$ at this stage. In brief, phonon bottleneck and Auger reheating effects greatly prolong the HC cooling in our PQDs. Specially, when $T_c$ cooled down to room temperature (~70 ps) the dynamic curve at IBP just decayed into zero, hence, the transient kinetic at IBP is demonstrated with more general applicability to evaluate the HC relaxation for different conditions than GSB formation, PIA decay and EAS.

**Hot carrier extraction dynamics**

HC extraction from $CsPbI_3$ QD to FDs is explored based on above discussion. Figure 2a displays the time-dependent carrier temperature $T_c$ of $CsPbI_3$ QD and $CsPbI_3$-FDs QD under same pump intensity (<N>~2.3). The photon numbers per QD are thought to be identical due to the negligible absorption of FDs, hence, extra loss of HCs is ascribed to the function of FDs. The TA signals of FDs in same concentration with $CsPbI_3$-FDs colloidal solution are shown in Figure S10, which are lower than the detection limit of our TA system. Surprisingly, CAFDs efficiently extract HCs from CsPbI$_3$ QDs proved by the obvious acceleration of $T_c$ decrease processes while CMFDs have insignificant impact on the HC (see more details about TA data in Figure S11-13). Since the temperature of HC has already shown apparent decrease around initial several picoseconds when cold carriers have not yet been extracted in quantity, it is reasonable to rule out the influence of cold carrier quenching on state filling and attribute the decrease of $T_c$ to the HC extraction. The carrier extraction induced electronic energy power loss rate ($J_e$) as a function of carrier temperature was further calculated to remove the influence of intrinsic HC cooling by the subtraction between the intrinsic $J_r$ and the $J_r$ for CsPbI$_3$-FDs QDs, that is $J_e(FDs)=J_r(CsPbI_3)-J_r(CsPbI_3\text{-}FDs)$. For all CAFDs, $J_e$ exhibits an apparent increasement starting from carrier temperature higher than 850K and increase steadily with $T_c$ (Figure 2b). Inset of Figure 2b indicates that HC transfer from CsPbI$_3$ QD boundary to CAFDs exists a threshold energy ($T_c$ > 850 K), which could be related to the location of energy level coupling (discuss latter). Figure 2c shows the transient kinetics at IBP for CsPbI$_3$ QDs with and without CAFDs (<N>~2.3). In consistent with temperature results, it is apparent that introduction of CAFDs accelerates the IBP decay (more IBP kinetics see Figure S14). By fitting the transient kinetics at IBP, the extraction efficiency ($\eta$) of HCs (Figure 2d) can be estimated according to $\eta=1-\tau_{CAFDs}/\tau_{QDs}$, where $\tau_{CAFDs}$ and $\tau_{QDs}$ are the lifetime of CsPbI$_3$-CAFDs QD and CsPbI$_3$ QD respectively. The results confirm that, improved HC lifetime due to bottleneck and Auger reheating under high pump fluence plays a crucial role in the HC harvesting. Long HC lifetime caused by these effects can keep HC above the threshold energy for longer time thus provides higher possibility to pass through the interface between PQD and selective materials. Finally, a maximum extraction efficiency (up to 84%) is acquired under <N>~2.3 by using [Bias]PCBA as selective materials. According to our knowledge, this result belongs to the premier echelon in HC extraction[30, 35-38]. We note that the small magnitude declines of $\eta$ under $<N>\sim5$ is possibly ascribed to the saturation of carriers in CAFDs under high pump power or other uncertain reasons.

**Molecule Interactions and electronic coupling**

The most striking observation to emerge from the data comparison were the entirely different HC extraction ability of CAFDs and CMFDs from $CsPbI_3$ QDs. Taking the molecule structures of PCBM and PCBA as comparison (Figure 3 a), the different function groups could be the driving force to their dramatically different HC extraction ability. In Figure 3b-c, analysis based on quencher's (FDs) concentration dependent fluorescence ($CsPbI_3$ QDs) quenching was adopted here to reveal the dynamic perspective of the molecular contact interaction hiding behind the following possible fluorescence quenching mechanisms[39]: (1) dynamic quenching, (2) static quenching (3) combined dynamic and static quenching (see detail illustration in supplementary note 3). Figure S15 shows the stable PL spectra of $CsPbI_3$ QDs at a known concentration of 0.83 mg/ml titrated with varying quantities of FDs. In order to acquire a reliable corresponding-relations between the PL quenching and the quantity of FDs, the purification process was canceled, instead, the mixture was straightly dissolved in chlorobenzene and stirred 5 mins. The quenching data are presented as a plot of $F_0/F$ versus $Q$ (Stern−Volmer plot, S-V plot, see details in Supplementary Note 4 about Stern−Volmer theory and the quenching models used in following) in Figure 3b-c, where $F_0$ and $F$ are the stable PL intensities in the absence and presence of FD quenchers, respectively, $Q$ is the concentration of FD quenchers. Surprisingly, CAFDs performed a more striking ability in quenching fluorescence than CMFDs; there is a hundred to one thousand times' difference at same quencher quantity. Besides, all the S-V plots appeared different types of derivation from the standard S-V equation that only exists one type of quenching (static or dynamic quenching) and presents a linearly dependence of fluorophores upon the concentration of quencher. A significant downward deviation from the linearity is displayed in the systems that use the CMFDs, i.e., PCBM and

[Bias]PCBM as quencher (Figure 3b). It is generally understood as dynamic quenching in presence of two populations of fluorophores that owns different inaccessibility to the quencher[40]. Possible explanation for this might be on account of the discrepancy in chemical circumstances on the surface of different QDs, such as atom terminal and ligand content[41]. Assuming the total PL ($F_0$) in the absence of quencher are contributed by inaccessible PQDs ($F_{i0}$) and accessible PQDs ($F_{a0}$), the S-V equation can be reframed as a linear form (inset of Figure 3b) according to $F_0/\Delta F = f_a^{-1} K_a^{-1} [Q]^{-1} + f_a^{-1}$, where $f_a$ is the fraction of initial PL contributed by accessible PQDs ($f_a = F_{a0}/F_0$), $K_a$ is quenching constant of accessible PQDs[40]. The fitting results are shown in Table 1. The higher quenching constant ($K_a$) and fraction ($f_a$) of accessible PQDs for [Bias]PCBM as quencher means more effective dynamic quenching than PCBM, which is probable due to double numbers of ester groups. For clarity, the upper image of Figure 3a use PCBM as the example to illustrate the above-mentioned dynamic interaction processes between the accessible PQDs and PCBM molecules. At dark state, the PCBM molecules and PQDs in colloidal solution keep independent with each other. When pump laser stimulates the PQDs from ground state to excited state, the transition dipole impels the PQDs to collide with PCBM and simultaneously quench excited carriers. After the quenching process, accessible PQDs and PCBM molecules separate each other and restore to initial state. The whole processes belong to physical interaction. Due to the existence of collision process in the solution, the carriers transfer is inefficient. This result is also supported by the unchanged ABS (Figure S6) and solution color (digital images in inset of Figure 3b) of $CsPbI_3$-CMFDs QDs compared with intrinsic $CsPbI_3$ QDs.

With regard to the quenching using CAFDs as quenchers, two stages are observed in both of the S-V plots : one is positive S-V plot, another is a linear S-V plot (see Figure 3c). Previous studies interpreted the positive deviation of S-V plot as either a combined dynamic and static (forming complex) quenching mechanism[42] or single static quenching but rather due to "sphere of action"[43]. Exactly distinguishing above two possibilities also needs to reconsider the ABS, which only reflects that ground state interactions. Essentially, the quenching within "sphere of action" is still achieved by the collision between quencher and excited state of fluorophores[39]. In contrast, the significant changes in the ABS take place only when CAFDs and $CsPbI_3$ QDs form ground state complex and part of their electronic state stably couple together. Based on these considerations, we propose to use the formula combining the dynamic quenching model and modified static quenching model to depict our positive S-V plots, that is $F_0/F=(K_D[Q+1])(K_B[Q]^n+1)$, where $K_D$ is dynamic quenching constant, $K_B$ is the binding constant of the static complex, $n$ is the number of equivalent binding sites for CAFDs onto per $CsPbI_3$ QD. From the fitting results in Table 1, $K_B$ is almost six orders of magnitude larger than $K_D$ indicating that the quenching by forming static complex holds the dominant position in the first quenching stage. The value n~2 obtained in all cases means about two CAFD molecules on average are bonding with per $CsPbI_3$ QD. The formation of static complex provides very high bonding stability and close contact distance from QDs to CAFDs, which facilitate the efficient coupling of electronic state for fast carrier transfer. The downer image of Figure 3a illustrates these interaction processes: firstly, the most PCBA molecules and PQDs have formed ground state complexes by chemical bonding in solution under dark state. Once the pump laser stimulates the complexes, fast carrier transfer occurs from $CsPbI_3$ QD to the bounding PCBA molecules. During this process, the collision induced slow carrier transfer also happens. Finally, after quenching, ground state complexes will contain their stable state while separate with the PCBA molecules that dynamically interact with QD. For $CsPbI_3$-CAFDs QDs, the core point lies on the formation of stable complexes that can be further demonstrated by the black dispersions under room light (digital images in inset of Figure 3c). X-ray photoelectron spectrum (XPS) and attenuated total reflection mode of Fourier transform infrared spectra (ATR-FTIR) illustrate that CAFDs would prefer chemically anchoring onto the uncapped atoms rather than extensively replacing the long chain ligand by traditional ligand exchange (See supplementary Figure S16-17 and Note 5), in agreement with the small number of equivalent binding sites. The linear S-V plot in the second stage indicates only one standard type of quenching occurs. Due to the binding sites have already been occupied, the specific interaction by forming static complex for excessive CAFD quenchers is nearly saturated within this stage. As result, it is easily attributing the quenching to standard dynamic quenching rather than standard static quenching although both display linear S-V plots. The equation describes this situation is $F_0/F=K_{D2}Q+b$, where $K_{D2}$ is dynamic quenching constant, b is constant[39]. The consistent orders of magnitude of $K_{D2}$ and $K_a$ (see table 1, PCBM vs PCBA,[Bias]PCBM vs [Bias] PCBA) also support the fact of dynamic quenching in this stage. Along with quenching stages, PL peak position also display related shift; detail discussions are shown in supplementary Note 6.

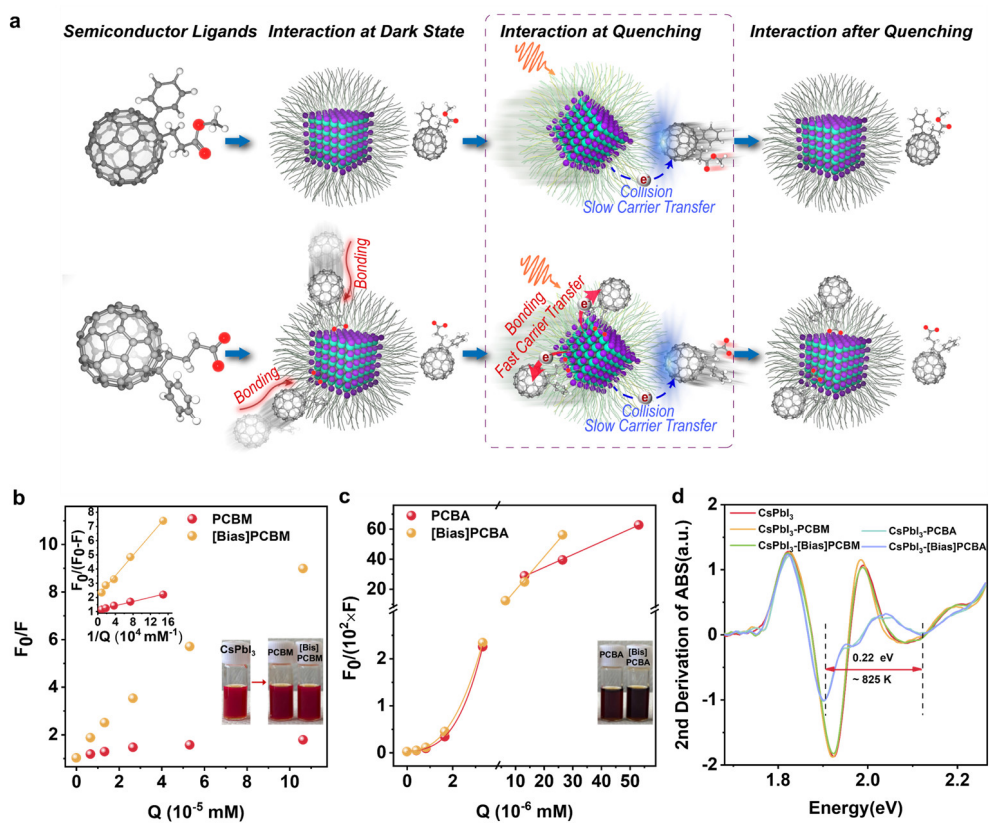

Figure 3. Physical and chemical interactions between $CsPbI_3$ QDs and FDs. (a) Schematic diagram of the interaction between PCBM/PCBA molecules and $CsPbI_3$ QDs. (b) CMFDs (i.e., PCBM and [Bias]PCBM), (c) CAFDs (i.e., PCBA

and [Bias]PCBA) quenching of CsPbI$_3$ QDs described in Stern−Volmer (S-V) plot. Inset of (b) is modified S-V plot to demonstrate the presence of two populations of fluorophores. (d) The second derivative absorption spectrum of CsPbI$_3$ QDs and CsPbI$_3$-FDs dispersions.

Table 1. Parameters fitted from S-V plots

|  | $K_a$ (M$^{-1}$) | $f_a$ | $K_D$ (M$^{-1}$) | n | $K_b$ (M$^{-n}$) | $K_{D2}$ (M$^{-1}$) |
|---|---|---|---|---|---|---|
| PCBM | 5.71×10$^7$ | 48.9% |  |  |  |  |
| [Bias]PCBM | 1.46×10$^8$ | 92.0% |  |  |  |  |
| PCBA |  |  | 2.47×10$^6$ | 2.06 | 2.09×10$^{12}$ | 8.52×10$^7$ |
| [Bias]PCBA |  |  | 1.21×10$^7$ | 2.12 | 1.84×10$^{13}$ | 2.22×10$^8$ |

Figure 3d shows the second derivative absorption spectra that eliminate the linear background function and further resolve the fine structure of electronic states. In quantum confinement systems, the minimal value points at second derivative ABS are thought to be related with the electronic state energy levels[36, 44]. Here, the weak absorption of FDs in the CsPbI$_3$-FDs dispersions has been removed before calculating derivative ABS. Thus, the changes are only associated with the effective coupling between FDs and CsPbI$_3$ QDs. From the area where significant differences locate at, it can be easily conjectured that the CAFDs have significant influences on the electronic states of CsPbI$_3$ QDs at both band edge and high energy level while CMFD cannot. For CsPbI$_3$-CAFD QDs, the energy difference between the minimal value points at high energy level (2.12 eV) and band edge (1.90 eV) is ~0.22 eV i.e., 825 K, in consistent with the HC transfer threshold, which implies the coupling electronic state provide channel for HC transferring from CsPbI$_3$ QDs to CDFA molecules. Besides, excitation power dependent PL spectrum (supplementary Figure S18) indicated the changes at band edge are attributed to electric-field induced exciton delocalization. This is also supported by the decrease of exciton binding energy (from 25 meV to 7 meV, clarified in supplementary Figure S19-20, Note 7 and Table S1) and highly efficient (~100%) cold electron transfer (Figure S21). The electric field induced exciton delocalization driven by FDs has also been reported in bulk perovskite film[45] and the theoretical calculations relying on the conception of the overall ligand/core adduct as an indecomposable species also demonstrated the mixing orbitals would lead to delocalized states throughout the whole band structure while QD and ligand experience each other's electric field[46, 47]. Thus, we infer that the electric-field tuned HC transfer may also occurred in our experiments.

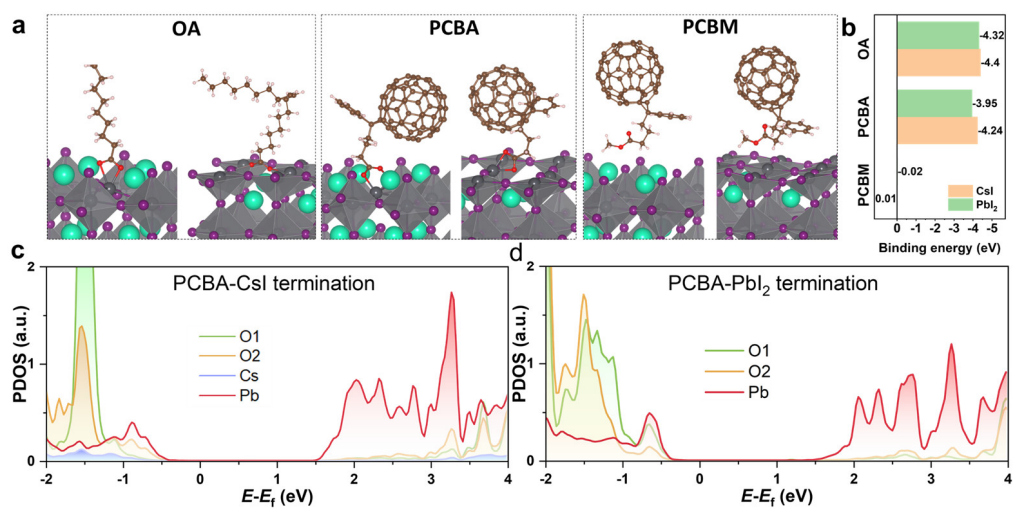

Figure 4. Density of function theory (DFT) calculations for QD-OA, QD-PCBA, and QD-PCBM, respectively. (a) The atomistic view of binding mode with CsI-termination (left figure) and PbI$_2$-termination (right figure), respectively. (b) The binding energies of ligands to the CsPbI$_3$ QDs. Partial density of state (pDOS) of the bonding atoms at the interface between PCBA molecule and (c) CsI termination, (d) PbI$_2$ termination.

**DFT calculation**

Density of function theory (DFT) was applied to further understand the nature of the HC extraction. Figure 4a shows the binding mode of OA, PCBA and PCBM with the optimized surface structures with CsI-termination and PbI$_2$-termination, respectively. More details of structural models and computational settings can be found in Supplementary Information and in Figure S22-23. Similar with OA, PCBA binds to both CsI- and PbI$_2$-terminated slabs via covalent bonds. On PbI$_2$-terminated QD, both Pb-O bond lengths ($L_{Pb-O(1)} \approx 2.51$ Å and $L_{Pb-O(2)} \approx 2.43$ Å) are smaller than the sum of Pb$^{2+}$ (1.2 Å) and O$^{2-}$ (1.4 Å) ion radii, illustrating that the effective bonding is realized by bidentate chelating between the two O atoms of PCBA and Pb atoms. Unlike $PbI_2$-termination, the bonding in CsI-termination is realized with the bidentate bridging mode, which have two O atoms bond with Pb atom and Cs atom separately. Whether the bonding mode is bidentate bridging or chelating, such bonding through multiple atoms has been shown to give extra stability to the complex than monodentate[48]. The binding energy (in Figure 4b) between PCBA and $CsPbI_3$ QD is large in negative values (about -4 eV) regardless of surface termination, illustrating PCBA can spontaneously anchor onto QD surface. In addition, slightly lower binding energy compared to OA (about -4.4 eV) suggests the slight preference of OA compared to PCBA, in agreement with experiments from S-V analysis, XPS and FTIR. In contract, the small binding energy (almost zero) between PCBM and $CsPbI_3$ QDs demonstrates the weak interaction. As shown in the charge density difference (CDD)(Figure S24), PCBM has no notable charge redistribution at the interface, indicating the PCBM molecular is physically adsorbed on the perovskite surface, which is different from the chemical bonding between PCBA and $CsPbI_3$ surface. From the partial density of state (pDOS) of the above-mentioned bonding atoms (Figure 4c-d), we see that large area of the synergic electronic state peak of bonding Pb atom with the electronic state of O atoms indicating their strong coupling. The high energy states coupling can be treated as the bridge for PQDs and selective materials, which extremely matters HC transfer at the interface. Besides, the overlap of the DOS between the whole PCBA with $CsPbI_3$ is another necessary condition for HC extraction (Figure S25). Taking OA as an example, despite the coupling at interface bonding (Figure S26), the carriers cannot continue transfer into OA molecule due to lacking further DOS overlap between the whole OA molecule with $CsPbI_3$ (Figure S25). In view of all the experimental and theorical results, we highlight the importance of the coupling of interface atoms, overlap of density state and interface electric field on HC extraction, in which the former helps interface carrier transfer between the coupling atoms, the middle decides on further extraction, the latter may accelerate whole processes.

## Conclusions

In this work, we explicate a general strategy to quantify HC extraction and achieve efficient extraction of HCs from $CsPbI_3$ QD to widely used organic FD acceptors with appropriately building QDs and FDs into state coupled complexes. For realizing extraction of HCs in FD and perovskite systems, the key is building their effective state couplings. We uncover that the stable chemical bonding in these complexes can acts as a bridge coupling FDs and $CsPbI_3$ QDs and provides effective channels for HC extraction. By adjust pump power, we finally achieve about 84% HC extraction efficiency. Our research provides an alluring strategy for utilizing HCs and pave the way to realize more efficient hot carrier photovoltaic technology.

## Author Contributions

Y.S. Li, Q. Shen conceived the idea. Y.S. Li conducted experiments, data analysis and paper writing. J.K. Jiang contributed to DFT calculations and the writing of the relevant results, S.X. Tao contributed to the supervision of DFT calculations, relevant data analysis and revision. D. Liu, Y. Shota, H. Li, F. Akihito, H.S. Li, G.Z. Shi, J.J. Shi and S. Hayase contributed to discussion. Q.B. Meng, C. Ding, D.D. Wang and Q. Shen contributed to discussion, data analysis and paper writing.

## Conflicts of interest

The authors declare no conflict of interest.

## Acknowledgements

The authors are very grateful to Prof. James Lloyd-Hughes (Department of Physics, University of Warwick, United Kingdom), Dr. Maurizio Monti (Institute for Physics and Astronomy, Denmark) for their kind providing of the TTM fitting code and patient explanations about their code. We also appreciate the meaningful help on FTIR analysis from Prof. Ishida Takayuki (The University of Electro-Communications, Japan). This work was supported by NEDO, the Japan Science and Technology Agency (JST) Mirai program (JPMJMI17EA), MEXT KAKENHI Grant (26286013, 17H02736), JST SPRING (Grant Number JPMJSP2131), the China Scholarship Council (No. 202008050282). S.T. and J.J. acknowledge funding by the Computational Sciences for Energy Research (CSER) tenure track program of Shell and NWO (Project number15CST04-2) and NWO START-UP, the Netherlands.

## Notes and references